# Magnetic tunnel junctions based on ferroelectric $Hf_{0.5}Zr_{0.5}O_2$ tunnel barriers

Yingfen Wei[1], Sylvia Matzen[2,*], Guillaume Agnus[2], Mart Salverda[1], Pavan Nukala[1], Thomas Maroutian[2], Qihong Chen[1], Jianting Ye[1], Philippe Lecoeur[2], Beatriz Noheda[1,*]

[1]Zernike Institute for Advanced Materials, University of Groningen, 9747 AG Groningen, The Netherlands

[2]Centre for Nanoscience and Nanotechnology, CNRS UMR 9001, Université Paris-Sud, Université Paris-Saclay, 91120 Palaiseau, France

* E-mail: sylvia.matzen@u-psud.fr, b.noheda@rug.nl;

## Abstract

A ferroelectric tunnel barrier in between two ferromagnetic electrodes (multiferroic tunnel junction, MFTJ), is one of the most promising concepts for future microelectronic devices. In parallel, Hafnia based ferroelectrics are showing great potential for device miniaturization down to the nanoscale. Here we utilize ferroelectric $Hf_{0.5}Zr_{0.5}O_2$ (HZO) with thickness of only 2 nm, epitaxially grown on $La_{0.7}Sr_{0.3}MnO_3$ (LSMO) ferromagnetic electrodes, as a large band-gap insulating barrier integrated in MFTJs with cobalt top electrodes. As previously reported for other MFTJs with similar electrodes, the tunneling magnetoresistance (TMR) can be tuned and its sign can even be reversed by the bias voltage across the junction. We demonstrate four non-volatile resistance states generated by magnetic and electric field switching with high reproducibility in this system.

**Introduction**

The concept of ferroelectric memory is by now a mature one[1]. The achievement of switchable ferroelectric polarization in ultra-thin films has opened possibilities for ferroelectric tunnel junctions (FTJs)[2–5]. Polarization switching of the ferroelectric barrier in a FTJs results in a change of the tunneling conductance, which is known as tunnel electroresistance (TER) effect. This phenomenon has been observed in several systems, such as $BaTiO_3$[6,7,8], $Pb(Zr_{0.2}Ti_{0.8})O_3$[9] and $PbTiO_3$[10]. Its origin has been mainly ascribed to three possible mechanisms[5]: a) charge screening at ferroelectric/electrode interfaces affecting the potential barrier profile; b) the change in the positions of ions at the interfaces after polarization reversal, or/and c) the strain differences induced by the electric field in the ferroelectric barrier.

Nevertheless, to achieve sufficiently thin ferroelectric films remains very challenging due to several issues, such as the difficulty to fully screen the surface polarization charges[11], the tendency of the films to form domains or other topological defects that cancel the net spontaneous polarization, the increase of the electric fields needed for polarization switching or the increase in the leakage currents. In the last few years, intensive research has been conducted on Hafnia-based thin films due to their unexpected ferroelectricity and to their CMOS compatibility.[12] Unlike all other known ferroelectrics, in Hafnia-based thin films, ferroelectricity becomes more robust as the size is decreased and it disappears above certain thickness in the range of 10-30 nm[13]. Thus, Hafnia-based thin films are highly promising as tunnel barriers for ferroelectric tunnel junctions[14]. Moreover, amorphous Hafnia is a high-k material that has been widely used as gate insulator in the microelectronic industry[15], so these thin films have great potential for applications in the next generation of memories and logic devices, showing great advantages compared to conventional perovskite ferroelectrics.

Multiferroic tunnel junctions (MFTJs), with a ferroelectric tunnel barrier integrated between two magnetic electrodes, instead of a linear-dielectric barrier (as in magnetic tunnel junctions, MTJs), were proposed a decade ago[16] and have become one of the most promising approaches to develop low-power, high-density, multifunctional and non-volatile, memory devices[17,18]. A MFTJ exhibits four non-volatile resistance states that can be achieved by external electric and magnetic field switching and are generated by the combination of the TER (determined by the ferroelectric polarization) and the TMR (tunnel magnetoresistance) effects. The latter originates in the dependence of the tunneling current on the parallel or antiparallel states between the two ferromagnetic electrode layers[19]. Previous studies on MFTJs have used ferroelectric tunnel barriers of $BaTiO_3$ or $PbTiO_3$/ $Pb(Zr,Ti)O_3$ (PZT), sandwiched between $La_{0.7}Sr_{0.3}MnO_3$ (LSMO) and Co magnetic electrodes[20–22].

Up to now, in Hafnia-based system, only amorphous, undoped and not ferroelectric layers have been used as tunnel barrier in magnetic tunnel junctions (MTJs)[23,24]. In our recent work, crystalline, rhombohedral $Hf_{0.5}Zr_{0.5}O_2$ (HZO) films have been grown epitaxially on (001)-LSMO (bottom electrode)/$SrTiO_3$ substrates and have shown ferroelectric switching with increasingly large remanent polarization values as the thickness decreases from 9 nm ($P_r$=18 µC/cm$^2$) down to 5 nm ($P_r$= 34 µC/cm$^2$).[25] Here, we report the first integration of ferroelectric HZO tunnel barriers in MFTJs, showing four non-volatile resistance states, as a combination of both TER and TMR effects.

**Results and discussion**

Ferroelectricity in 2 nm thick HZO films

Thin layers of HZO with thickness of 2 nm were grown on LSMO-buffered STO substrates. As shown by Piezoresponse Force Microscopy (PFM) images in Fig. 1a-d, the films display all the characteristics of ferroelectric behavior. Fig.1a shows the phase image, after alternatively applying +7 V and -7 V biases on the back electrode in successively smaller sample square areas. The large observed contrast is consistent with the electrical polarization of the films pointing in opposite directions for the different polarities of the applied field. No influence of this electrical writing has been observed in the amplitude (Fig. 1b) or topography (Fig. 1c) images. Furthermore, by Kelvin Probe Force Microscopy (KPFM), as seen in Fig. 1d, the surface potential difference between the area written by 7 V and -7 V is found to be around 600 mV, a value comparable to other conventional ferroelectrics, e.g. $BaTiO_3$[26], PZT films[27,28]. Both phase and potential contrasts are stable for at least 48 h.

The 2 nm thick HZO ferroelectric ultrathin films characterized above have been integrated with top Co magnetic electrodes to fabricate MFTJs with areas ranging from 10 x 10 µm$^2$ to 30 x 30 µm$^2$ (see methods section for details). The cross-section Scanning Transmission Electron Microscopy (HAADF-STEM) image presented in Fig. 1e shows sharp interfaces between LSMO / rhombohedral (111)-oriented HZO layers[25] and polycrystalline Co. The schematic view of a complete MFTJ device is shown in Fig. 1f.

HZO-based magnetic tunnel junction (MTJ)

The current-voltage (I-V) characteristics of 2 nm- and 3 nm-thick films with the same junction area (20 x 20 µm$^2$) are shown in Fig. 2a. Current through the 3 nm-thick HZO film is too low (below 1 nA) to be reliably measured with our experimental setup and a thinner film is required for a tunneling junction. Indeed, the parabolic dependence of the differential conductance of the 2 nm film, which is fitted using the Brinkman model[29], indicates that the transport mechanism is direct tunneling through the HZO barrier. Due to the large band gap (5-6 eV) of

HZO, the junction is very resistive even for ultrathin films thus preventing break-down problems and improving the stability of the devices. All further measurements are performed on different devices with the same ultrathin 2 nm-thick barrier.

The magnetic hysteresis loop M(H) of a similar (but unpatterned) sample at 50 K is shown in Fig. 2b, with the magnetic field applied along the in-plane [110] easy axis direction of LSMO. The magnetic switching of both LSMO and Co layers is clearly observed, showing coercive fields of around +/- 50 Oe for LSMO and +/- 250 Oe for Co. This difference allows for an antiparallel magnetic alignment between both magnetic electrodes for intermediate magnetic fields.

The resistance of such devices is measured as a function of magnetic field under a bias of -0.2 V (applied to the top Co electrode) at a temperature of 50 K in a 10 x 10 µm² junction, for magnetic field cycling from 2000 Oe to -2000 Oe and back, along the [110] axis (Fig. 2c). A higher resistance state is measured in antiparallel magnetic configuration when sweeping the field, displaying a positive TMR value of 5.4%, where TMR is defined as $(R_{AP}-R_p)/R_p$, with $R_{AP}$ and $R_p$ the resistance values in antiparallel and parallel states, respectively. This low value is in line with the expected low spin polarization of the 3d ferromagnetic metal. The record TMR among the Co-based MFTJs is around 30% for an epitaxial single crystalline $PbTiO_3$ barrier[20]. In addition, TMR effect disappears for temperatures above 250 K (Supplementary Fig. S1), in agreement with most studies performed on other MFTJs with LSMO and Co electrodes.[22] While the Curie temperature of LSMO magnetic electrode is found to be around 350 K (Supplementary Fig. S2), the disappearance of TMR at lower temperatures could be a result of either the decrease of the spin polarization of LSMO at the interface with HZO, and/or the spin-independent tunneling through impurity levels in the barrier activated upon increasing the temperature.[30–34]

Junctions with different sizes have been fabricated, and six of them with a STO/LSMO/HZO (2nm)/Co stack were connected to a chip carrier and measured. They all show TMR ratios between 5% and 7% under -0.2 V bias at the temperature of 50 K (Fig. 2d). This high reproducibility in the properties of junctions proves the excellent quality of the HZO tunnel barrier, despite the domain-like nanostructure of the films[25]. In addition, the resistance-area product (RA) is greater than 600 MΩ·µm² for all junctions shown in Fig.2d, which is orders of magnitude higher than typically encountered in MFTJs with perovskite ferroelectric layers[22], again highlighting the performance of ultrathin HZO as tunnel barrier.

## Four resistance states

In the present case of the HZO barrier, we observe a resistance switching behavior as seen in previous works on conventional perovskite ferroelectric barriers[6–9,35]. A resistance hysteresis loop indicating memristive behavior is shown in Fig. 3a. The junction resistance measured under a bias of 0.1 V is plotted as a function of the amplitude of the successive write pulses (500 μs pulse width). A clear hysteresis cycle between a low ($R_{ON}$) and high ($R_{OFF}$) resistance state is achieved, with ON/OFF ratio of ~440%, defined as $R_{OFF}/R_{ON}$. The switching voltage between both states is around 2 V, when the write pulse is swept from -6 V to 6 V, and around -2 V when going back to -6 V. The low resistance state ($R_{ON}$) corresponds to the ferroelectric polarization up ($P_↑$), which is the as-grown state of the HZO films, as indicated in Fig. 1a. The highest resistance state is observed when the polarization is pointing towards the bottom LSMO electrode, as also reported in PZT [9] and BTO[7,35] based MFTJs with LSMO and Co electrodes.

In Fig. 3b, TMR loops are obtained after +6 V ($R_{OFF}$), and -6 V ($R_{ON}$) pulses and show both TMR ratio of around 5.2 %, corresponding to TER~ 190%. Four resistance states can thus be obtained, and switched reversibly either electrically or magnetically. One can observe that the TMR does not change significantly between ON and OFF states. The spin polarization of the tunneling electrons thus appears unaffected by the ferroelectric switching. In addition, the TMR loops at different resistance states (after pulses of -6 V and 6 V, respectively) for a different device are also shown in Fig. S3, confirming that TMR displays only small changes for different electric field polarities. Furthermore, to study the dependence of TMR with bias, I(V) curves are measured in both parallel and antiparallel states. From these measurements, the TMR ratio can be extracted at different bias since TMR=$(I_P-I_{AP})/I_{AP}$, where $I_{AP}$ and $I_P$ are the current in antiparallel and parallel states, respectively. Fig. 3c shows that the TMR ratio bias-dependence is barely affected by the ferroelectric polarization state. This proves once again the stability of the resistance states, but also the absence of measurable magnetoelectric coupling[21,23] in this system.

## Inverse TMR

As shown in Fig. 4a, when a positive bias of 0.2 V is applied on the top electrode Co, an inverse TMR (of around -2.6%) is observed at 50 K, corresponding to a smaller resistance measured in the antiparallel state compared to the parallel one. From the resulting TMR(V) curve (red) in Fig. 4b at the same temperature (see also Fig. 3c), the largest TMR (~ 6%) is measured at a bias of about -0.3 V. The inverse TMR can be observed above a threshold bias value around 0.1V at this temperature.

According to Jullière's model[36], the TMR amplitude and sign are related to the spin polarization of the density of states (DOS) of the two ferromagnetic layers. In particular, for the case of tunneling between LSMO and Co electrodes, applying different bias changes the relative position of the DOS of Co and LSMO, as depicted by De Teresa *et al.*[37] for a SrTiO$_3$ barrier. In their case, with the bias applied on the back LSMO electrode, the negative TMR is observed when bias is below ~0.8 V at 5 K.

Moreover, in the case of the HZO barrier, TMR(V) curves are also plotted in Fig. 4b at different temperatures. The bias at which the TMR sign changes is defined as $V_{TMRsign}$. Interestingly, we observe that $V_{TMRsign}$ increases with temperature, from ~0.1 V at 20 K to ~0.35 V at 200 K, as shown in Fig. 4c (in blue). This could be due to the decreasing spin polarization of LSMO at the interface with HZO with increasing temperature, as the decrease of TMR shows a similar trend (plotted in black in Fig. 4c with values extracted from Fig. S1).

## Conclusion

We have successfully built the first MFTJs with ultra-thin ferroelectric Hafnia-based barrier showing high quality and stability, and we have realized four non-volatile states. The junctions display all the expected characteristics, such as: 1) bias-dependent inverse TMR, originated by the LSMO and Co electrodes combination; 2) an increase in the voltage needed for TMR sign change ($V_{TMRsign}$) that increases with increasing temperature, following the decrease in the TMR values; 3) memristive behavior, as expected for ferroelectric barrier. All of the above shows the great potential of this material for multifunctional devices and adaptable electronics.

## Methods

**Thin-film synthesis:** Thin films of $Hf_{0.5}Zr_{0.5}O_2$ (HZO) barrier with thickness of 2 nm were grown by pulsed laser deposition (PLD) on ferromagnetic (FM) $La_{0.7}Sr_{0.3}MnO_3$(LSMO)-buffered (001)-$SrTiO_3$ substrates. Thickness of LSMO film is 30 nm. Details of the growth conditions can be found in Ref.[25]. 50 nm FM Cobalt with a protective layer of Au (50nm), to preserve Co from oxidation, was deposited by sputtering on top of HZO, to form LSMO (FM) / HZO (FE)/ Co (FM) stack.

**PFM and KPFM analysis:** Piezoelectric force microscopy (PFM) and Kelvin Probe force microscopy (KPFM) analysis was performed using an Asylum Cypher Atomic Force Microscope (AFM) with two different tip types: Rocky Mountains 25PtIr300B tips for writing and Adama AD-2.8-AS tips for reading. For writing, a bias of +/- 7 V was applied to the back electrode, while the tip was grounded. For reading the PFM signal, the AFM was operated in DART mode with the PFM drive signal on the tip and with a bias of ~0.6 V applied to the back electrode to reduce the electrostatic interactions between the sample and the AFM cantilever. This offset bias corresponds to the surface potential difference between the written up and down domains, measured by KPFM using the same tip.

**Device fabrication:** Junctions of different sizes (10 x 10 µm$^2$, 20 x 20 µm$^2$, 30 x 30 µm$^2$) were fabricated by photolithography, chemically assisted ion beam etching (IBE) controlled by a secondary ion mass spectrometer (SIMS), and sputtering of metallic top electrodes and $Si_3N_4$ insulating layer in different steps. The schematic drawing and cross-section TEM picture of device are shown in Fig. 1e and f. Junctions are connected by wire-bonding to chip carrier. The low temperature and magnetic field are applied in a Physical Properties Measurement System (PPMS) by Quantum Design. The electrical measurements are performed using a Keithley 237 source measurement unit, and the electrical pulses are done with a Keithley 4200A-SCS parameter analyzer.

Figures:

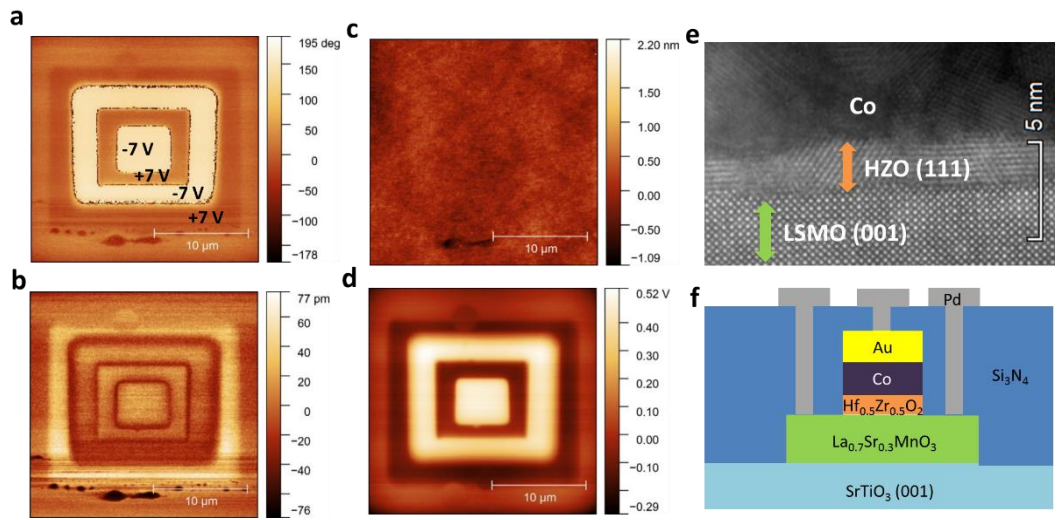

**Fig. 1 Ferroelectricity in 2 nm thick HZO films. a,** Piezoresponse (phase) measured after polarization switching of a 2 nm HZO layer with +7 V and -7 V alternative writing voltages applied to the bottom LSMO electrode. **b,** PFM out-of-plane amplitude and **c,** AFM topography (25 x 25 μm$^2$) in the same region as phase in **a**. **d,** Surface potential measured by KPFM. **e,** HAADF-STEM image of cross-section LMSO/HZO/Co stack. **f,** Schematic drawing of junction device.

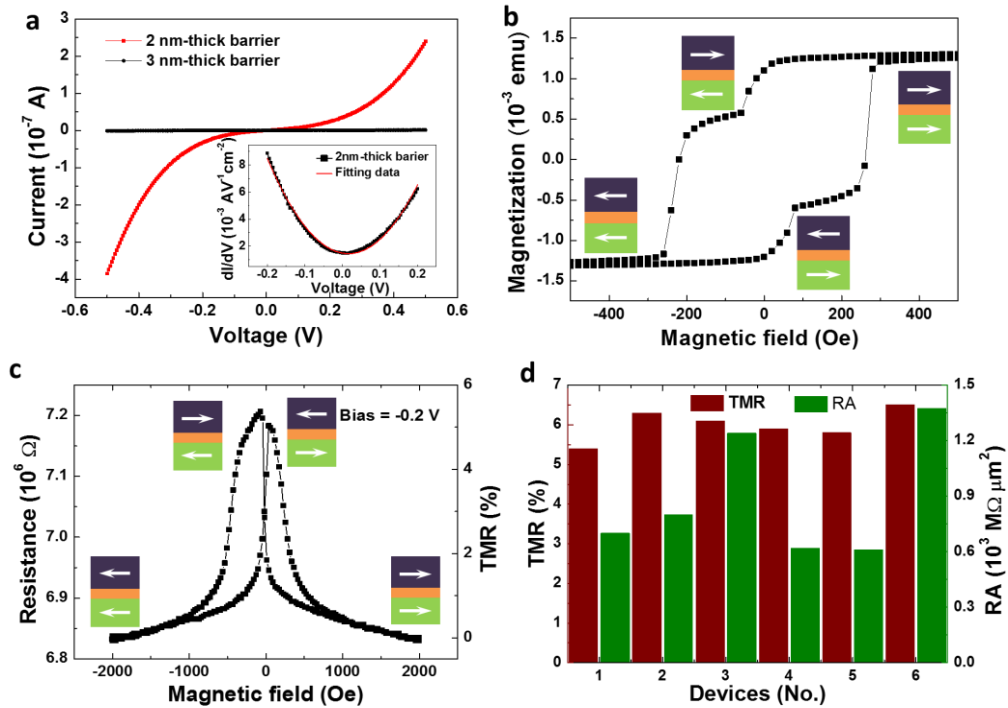

**Fig. 2 HZO-based MTJs. a,** I(V) curves at 50 K of 20 x 20 µm² junctions with 2 nm- and 3 nm-thick barriers. Inset shows the derivative of the I-V curve for the 2 nm film, with the parabolic Brinkman fit. **b,** M(H) loop of an unpatterned sample measured at 50 K by superconducting quantum interference device (SQUID) magnetometry along the in-plane [110] direction of LSMO. **c,** TMR loop measured in junction 1 (10 x 10 µm²) under bias of -0.2 V at 50 K, with high (low) resistance in antiparallel (parallel) state. **d,** TMR and resistance area product (RA) in all the measured junctions on the same sample with 2 nm thick HZO barrier.

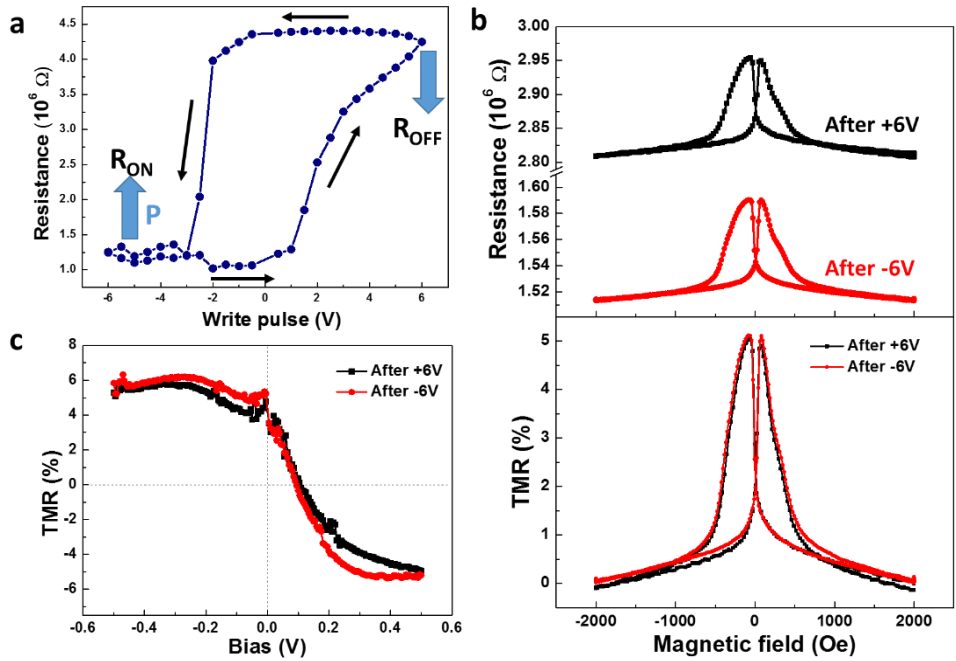

**Fig. 3 Combined TMR and TER. a,** Resistance hysteresis loop (read by a voltage of 100 mV) as a function of write pulses with different amplitudes from -6 V to +6 V and width of 500 µs on junction 6 (30 x 30 µm$^2$). Blue arrows indicate the orientation of the ferroelectric polarization as up (P$_\uparrow$, towards the Co electrode) and down (P$_\downarrow$, towards the LSMO electrode). **b,** Resistance as a function of magnetic field (upper panel), and corresponding TMR loops (lower panel) under a bias of -0.2 V at 50 K, and **c,** bias-dependent TMR ratio after +6 V (P$_\downarrow$) and -6 V (P$_\uparrow$) pulses on junction 2 (20 x 20 µm$^2$).

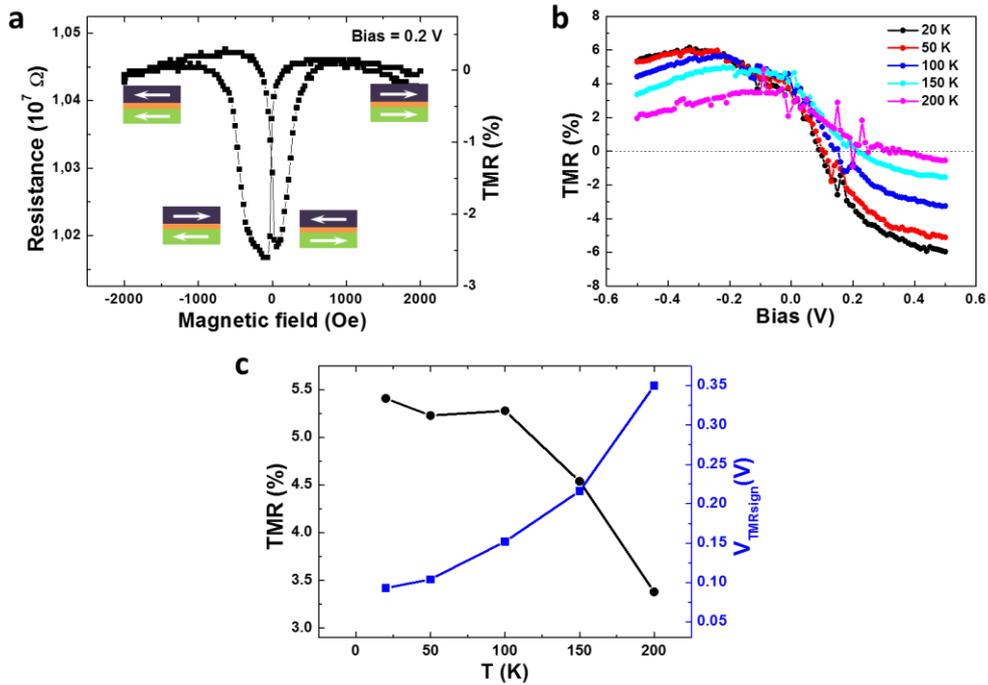

**Fig. 4 Inverse TMR. a,** TMR loop obtained in junction 1 (10 x 10 µm$^2$) under a bias of 0.2 V at 50 K with high (low) resistance in parallel (antiparallel) state. **b,** Bias-dependent TMR from -0.5 V to 0.5 V at different temperatures from 20 K to 200 K. **c,** Temperature dependence of both TMR (black, circles) and V$_{TMRsign}$ the voltage needed for TMR sign reversal (blue, squares) in the same junction.

**Supplementary figures:**

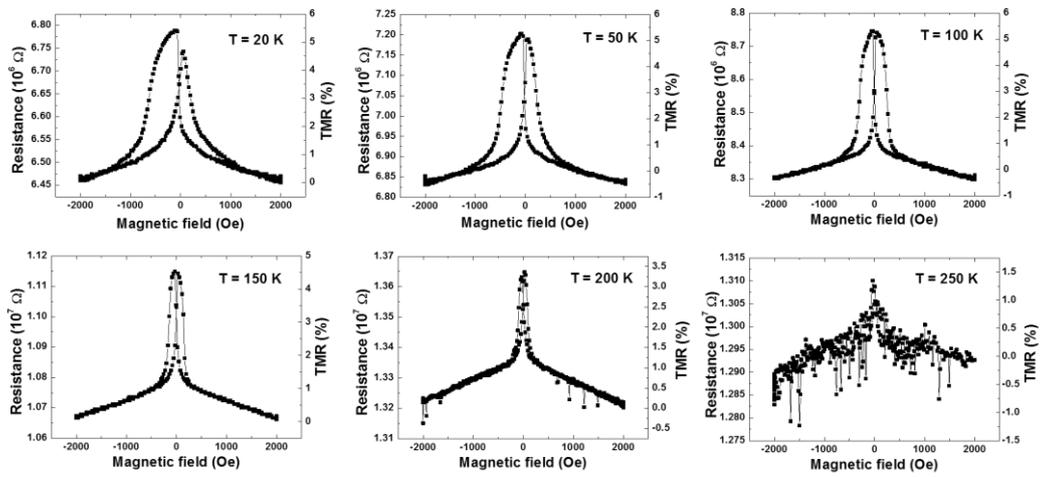

**Fig. S1** TMR ratios of junction 1 (10 x 10 µm$^2$) under bias of -0.2 V at different temperatures from 20 K to 250 K.

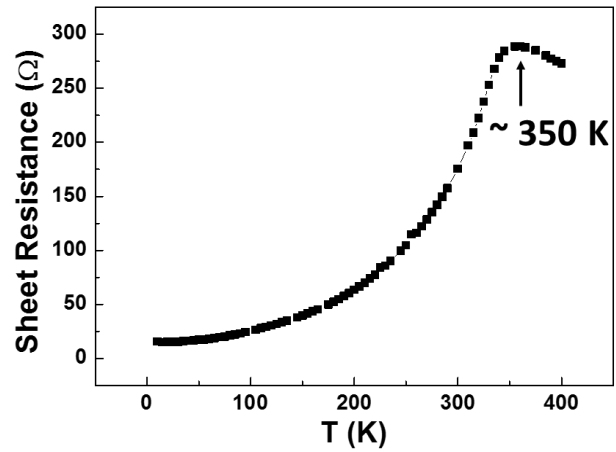

**Fig. S2** Sheet resistance of STO/LSMO as a function of temperature showing the transition of the LSMO layer.

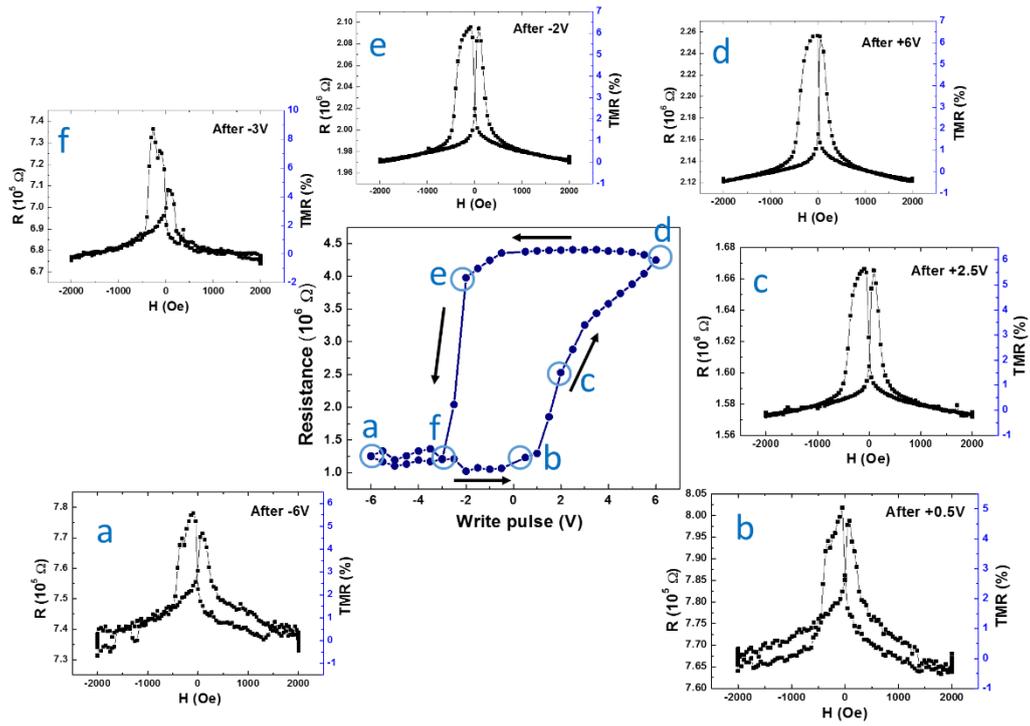

**Fig. S3** TMR loops under bias of -0.2 V at different resistance states and 50 K on junction 6 (30 x 30 µm$^2$).